# Binary Random Sequences Obtained From Decimal Sequences

Suresh B. Thippireddy
Oklahoma State University, Stillwater

**Abstract:** This paper presents a twist to the generation of binary random sequences by starting with decimal sequences. Rather than representing the prime reciprocal sequence directly in base 2, we first right the prime reciprocal in base 10 and then convert it into the binary form. The autocorrelation and cross-correlation properties of these binary random (BRD) sequences are discussed.

## Introduction

Random numbers, which are central to many scientific and cryptographic applications, may be generated using different approaches, ranging from classical to quantum. Classically generated numbers (as in [14], [15]) have their randomness predicated on complexity; quantum random number generators exploit the fundamental randomness of quantum phenomena.

D sequences [1-15] are perhaps the simplest family of random sequences that subsumes other families such as shift register sequences [16]. In their ordinary form, d sequences are not computationally complex [2], but they can be used in a recursive form [8] that is much stronger from a complexity point of view. The basic method of the generating the binary d-sequences is given in [1]. The autocorrelation properties of the d sequence based recursive random number generator were studied in [10] where an approach for multi-recursive random number generator was also proposed. These methods were all based on a direct binary representation of appropriately chosen numbers.

In this note we take a different approach to the generation of binary d sequences by beginning with a base 10 (decimal) sequence first and then converting each digit to a binary form. The advantage of this is that longer period binary sequences are now obtained. For example, the straight binary representation of 1/7 has a period of 3 (the sequence being 001), whereas if we begin with the base-10 expansion of 1/7 which is 142857 and then convert it into the binary form digit by digit (using shortest representations) we get

     1100101000101111

The above sequence is 15 bits long.

If equal length representations of the digits are used, we get a binary sequence that is 24 bits long.

     000101000010100001010111

Several properties and results of this approach are shown in this paper.

## Method:

According to the standard method, the binary d-sequence is generated using the algorithm [1-3] below:

$$a(i) = 2^i \bmod p \bmod 2$$

where p is a prime number. The maximum length period (p-1) sequences are generated when 2 is a primitive root of p.

It is worthwhile to determine if this phenomenon holds for non-binary cases. To generate decimal expansions of 1/p, one may use the following formula [13]:

If prime ends in 1,
$$a(i) = 9 \times 10^i \bmod p \bmod 10$$

If prime ends in 3,

$$a(i) = 3 \times 10^i \bmod p \bmod 10$$

If prime ends in 7,

$$a(i) = 7 \times 10^i \bmod p \bmod 10$$

If prime ends in 9,

$$a(i) = 10^i \bmod p \bmod 10$$

The above formula produces a random sequence of decimal numbers ranging from 0 to 9. These decimals are then converted to an equivalent binary numbers, which results in a sequence of binary numbers.

In this paper, we consider the equal length representations for converting the decimal values to the equivalent binary numbers.

## Autocorrelation properties

Having the sequence generated using the above procedure; we then look for the autocorrelation properties to determine how good they are from the point of view of randomness. The autocorrelation function is calculated using the formula:

$$R(k) = \frac{1}{n}\sum_{j=0}^{n-1} a_j a_{j+k}$$

$a_j$, $a_{j+k}$ are the binary values of the sequence generated by the above process, and n is the length of the sequence.

**Examples:**

1) For p=17,

    The decimal sequence generated using the modulo 10 method is:
    0588235294117647
    The length of decimal sequence is 16

    The binary sequence equivalent to the above decimal sequence is:
    0000010110001000001000110101001010010100000100010111011001000111
    The period of the binary sequence is 64

    The auto-correlation values of the above binary sequence are in the below graph:

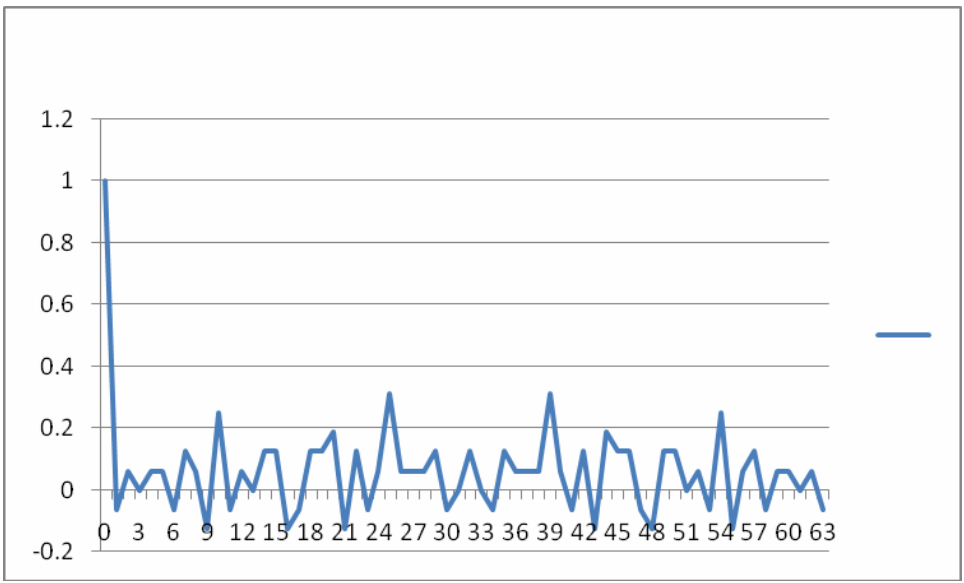

Figure 1. Autocorrelation for binary sequence generated using p=17; period = 64

2) For p = 149,
   The length of decimal sequence is 148
   The period of the binary sequence is 592
   The auto-correlation values of the above binary sequence are in the below graph:

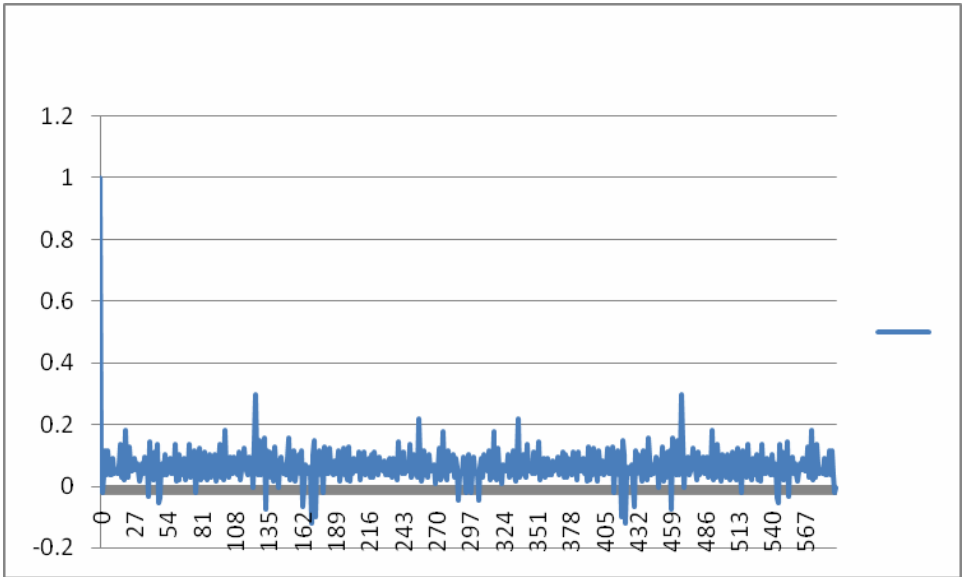

Figure 2. Autocorrelation for binary sequence generated using p=149; period = 592

3) For p = 457,
   The length of decimal sequence is 152
   The period of the binary sequence is 608
   The auto-correlation values of the above binary sequence are in the below graph:

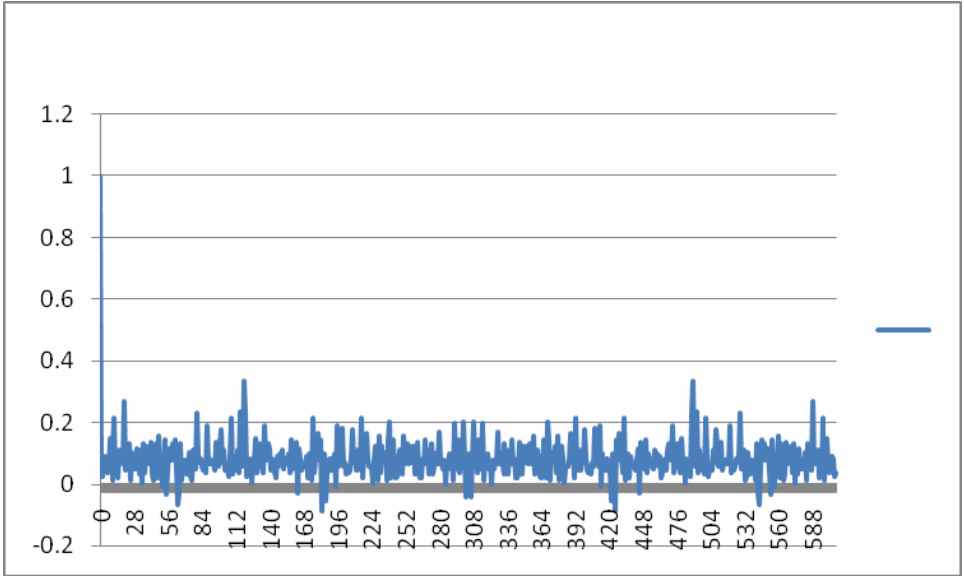

Figure 3. Autocorrelation for binary sequence generated using p=457; period = 608

## Cross-correlation properties:

The cross-correlation of two sequences is calculated using the below formula:

$$C(k) = \frac{1}{q}\sum_{j=0}^{q-1} a_j b_{j+k}$$

$a_j$ represent the binary value of the sequence a; $b_{j+k}$ represents the binary value of the sequence of b; q is the length of the sequences, it is calculated as below:

$$q = \text{LCM}(n1, n2)$$

n1 is the period of the sequence 'a'. n2 is the period of the sequence 'b'.

## Examples:

1) For primes 17 and 19,
   Period of the sequences for the primes 17 and 19 are n1=64, n2=72.

We know that in order calculate the cross correlation of the above two primes, we need calculate the value of q=LCM (n1, n2).

Here we get q=576

The cross-correlation values for the two primes 17, 19 are shown in below graph:

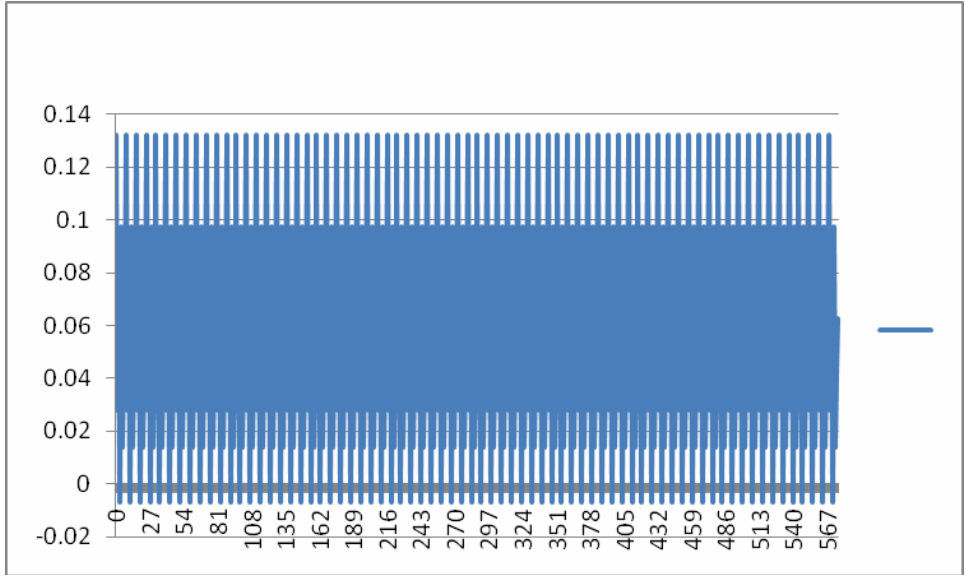

Figure 4. Cross-correlation for binary sequences generated using p=17, 19

2) For primes 137 and 257,
   Period of the sequences for the primes 137 and 257 are n1=32, n2=1024.
   We know that in order calculate the cross correlation of the above two primes, we need calculate the value of q=LCM (n1, n2). Here we get q=1024

   The cross-correlation values for the two primes 137, 257 are shown in below graph:

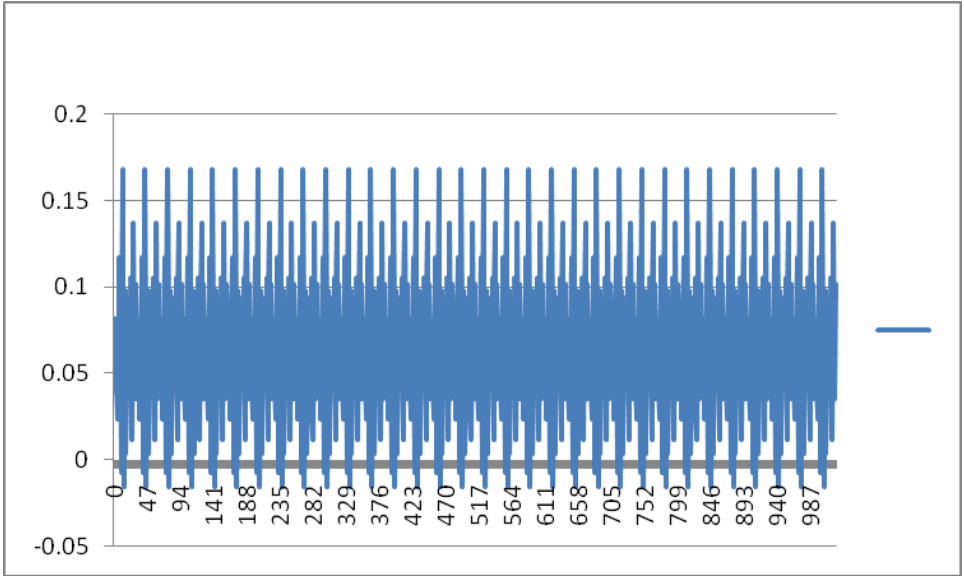

Figure 5. Cross-correlation for binary sequences generated using p=137, 257

The effectiveness of random sequences in many multi-access situations is determined by how small the maximum of the cross-correlation function is. Table 1 presents the cross-correlation values for several small primes and Table 2 presents the maximum of these values.

**Table 1: Maximum cross-correlation values for different primes**

| Primes | 17 | 31 | 53 | 89 | 113 | 137 | 151 | 199 | 257 | 283 | 331 |
|---|---|---|---|---|---|---|---|---|---|---|---|
| 17 | 1 | 0.137 | 0.086 | 0.130 | 0.191 | 0.312 | 0.123 | 0.125 | 0.160 | 0.111 | 0.125 |
| 31 | 0.137 | 1 | 0.125 | 0.137 | 0.136 | 0.133 | 0.26 | 0.173 | 0.136 | 0.143 | 0.165 |
| 53 | 0.086 | 0.125 | 1 | 0.095 | 0.087 | 0.096 | 0.106 | 0.103 | 0.088 | 0.093 | 0.080 |
| 89 | 0.130 | 0.137 | 0.095 | 1 | 0.123 | 0.136 | 0.123 | 0.162 | 0.117 | 0.113 | 0.161 |
| 113 | 0.191 | 0.136 | 0.087 | 0.123 | 1 | 0.174 | 0.121 | 0.122 | 0.136 | 0.111 | 0.011 |
| 137 | 0.312 | 0.133 | 0.096 | 0.136 | 0.174 | 1 | 0.116 | 0.113 | 0.167 | 0.111 | 0.095 |

| | | | | | | | | | | |
|---|---|---|---|---|---|---|---|---|---|---|
| 151 | 0.123 | 0.26 | 0.106 | 0.123 | 0.121 | 0.116 | 1 | 0.141 | 0.122 | 0.133 | 0.131 |
| 199 | 0.125 | 0.173 | 0.103 | 0.162 | 0.122 | 0.113 | 0.141 | 1 | 0.122 | 0.125 | 0.154 |
| 257 | 0.160 | 0.136 | 0.088 | 0.117 | 0.136 | 0.167 | 0.122 | 0.122 | 1 | 0.111 | 0.114 |
| 283 | 0.111 | 0.143 | 0.093 | 0.113 | 0.111 | 0.111 | 0.133 | 0.125 | 0.111 | 1 | 0.109 |
| 331 | 0.125 | 0.165 | 0.080 | 0.161 | 0.011 | 0.095 | 0.131 | 0.154 | 0.114 | 0.109 | 1 |

The minimum of the maximum cross-correlation value for the cases we have considered is 8 percent.

**Table 2: Minimum cross-correlation values for different primes**

| primes | 17 | 31 | 53 | 89 | 113 | 137 | 151 | 199 | 257 | 283 | 331 |
|---|---|---|---|---|---|---|---|---|---|---|---|
| 17 | -0.12 | 0.070 | 0.067 | 0.005 | -0.02 | -0.25 | 0.052 | 0.049 | -0.02 | 0.027 | 0.009 |
| 31 | 0.070 | 0 | 0.1 | 0.075 | 0.069 | 0.05 | 0.006 | 0.070 | 0.070 | 0.058 | 0.013 |
| 53 | 0.067 | 0.1 | -0.15 | 0.079 | 0.063 | 0.057 | 0.082 | 0.079 | 0.065 | 0.063 | 0.056 |
| 89 | 0.005 | 0.075 | 0.079 | -0.09 | 0.030 | -0.04 | 0.058 | 0.023 | 0.035 | 0.048 | -0.02 |
| 113 | -0.02 | 0.069 | 0.063 | 0.030 | -0.12 | -0.04 | 0.056 | 0.048 | 0.026 | 0.032 | 0.014 |

| 137 | -0.25 | 0.05 | 0.057 | -0.04 | -0.04 | -0.25 | 0.036 | 0.035 | -0.01 | 0.015 | 0.015 |
| 151 | 0.052 | 0.006 | 0.082 | 0.058 | 0.056 | 0.036 | -0.01 | 0.067 | 0.056 | 0.045 | 0.034 |
| 199 | 0.049 | 0.070 | 0.079 | 0.023 | 0.048 | 0.035 | 0.067 | -0.06 | 0.048 | 0.041 | 0.016 |
| 257 | -0.02 | 0.070 | 0.065 | 0.035 | 0.026 | -0.01 | 0.056 | 0.048 | -0.10 | 0.034 | 0.017 |
| 283 | 0.027 | 0.058 | 0.063 | 0.048 | 0.032 | 0.015 | 0.045 | 0.041 | 0.034 | -0.14 | 0.012 |
| 331 | 0.009 | 0.013 | 0.056 | -0.02 | 0.014 | 0.015 | 0.034 | 0.016 | 0.017 | 0.012 | -0.05 |

The maximum of the minimum of the cross-correlation value for the cases considered here is 8.2 percent.

## Conclusions

The autocorrelation and cross-correlation of the binary random sequences generated using a decimal sequence generator is investigated, and it is shown that their properties are excellent. There are BRD pairs for which the maximum cross-correlation is less than 8 percent. This is much better than the performance of PN sequences [17].

The results in this note were found by mapping the digits in the decimal sequence using equal 4-bit representation. One can also investigate properties where unequal mapping of the digits is chosen.